\documentclass[10pt,letterpaper]{article}

\usepackage{amsmath,amssymb}
\usepackage{amsmath}
\usepackage{graphicx}
\usepackage{booktabs}
\usepackage{longtable}
\usepackage{array}
\usepackage{multirow}
\usepackage{caption}
\usepackage{subcaption}
\usepackage{setspace}
\usepackage{lineno}
\usepackage{authblk}
\usepackage[hidelinks]{hyperref}
\usepackage{tikz}
\usetikzlibrary{positioning, arrows.meta, calc, shapes.geometric}

\title{A minimal E/I circuit model explains differential recovery from amblyopia under TTX and reverse occlusion}

\author[1]{Lawrence Udeigwe}
\affil[1]{Department of Mathematics, Manhattan University, Riverdale, NY, USA}

\date{}

\begin{document}

\maketitle

\begin{abstract}
Recovery from amblyopia beyond the critical period remains a major challenge in visual neuroscience. Experimental studies have shown that transient inactivation of the fellow eye using tetrodotoxin (TTX) produces stronger and more persistent recovery than reverse occlusion (RO), despite the fact that both interventions reduce visual competition from the fellow eye. The circuit mechanisms underlying this difference remain unclear.

Here, we develop a minimal cortical circuit model that links recovery from amblyopia to the dynamical state of the cortical network. The model combines excitatory--inhibitory population dynamics, BCM-like metaplasticity, and homeostatic regulation within a two-population circuit receiving eye-specific inputs. Plastic feedforward synapses mediate competition between amblyopic and fellow-eye pathways, while adaptive modification thresholds regulate activity-dependent plasticity.

Monocular deprivation drives the network into an amblyopic operating regime characterized by reduced excitatory activity and fellow-eye dominance. Phase-plane analysis and operating-regime characterization reveal that TTX and RO subsequently move the circuit into distinct dynamical states. TTX produces a larger plasticity-permissive regime by reducing inhibitory activity while maintaining excitatory activity above the BCM modification threshold. In contrast, RO preserves greater inhibitory tone and generates a smaller plasticity-permissive window. These dynamical differences lead to stronger recovery of amblyopic-eye responsiveness and greater correction of ocular dominance under TTX. Robustness analysis demonstrates that this recovery advantage persists across a broad range of parameter perturbations.

Together, these results provide a mechanistic explanation for the superior efficacy of fellow-eye inactivation and suggest that recovery from sensory deprivation is governed not only by changes in sensory input, but also by the dynamical operating regime of the cortical circuit.
\end{abstract}

\section*{Author Summary}
Amblyopia is a developmental visual disorder in which abnormal early visual experience weakens cortical responses associated with one eye. Traditional treatments become less effective after the critical period, making recovery in the mature brain particularly challenging. Recent experiments have shown that temporarily silencing the stronger eye with tetrodotoxin (TTX) can produce substantial recovery, often exceeding that achieved through reverse occlusion (RO). However, the reason for this difference has remained unclear.

We developed a simple computational model of cortical plasticity to investigate how these interventions affect recovery. The model combines excitatory--inhibitory circuit dynamics with BCM-like metaplasticity and homeostatic regulation. Although both TTX and RO reduce competition from the fellow eye, the model predicts that they place the cortex into different dynamical operating regimes. TTX produces a stronger reduction in inhibition and creates a larger plasticity-permissive window, allowing deprived-eye pathways to recover more effectively. RO generates a weaker shift in circuit state and consequently produces less recovery.

Our results suggest that successful recovery depends not only on sensory input itself, but also on the dynamical state of the cortical circuit through which that input is processed. More broadly, the work highlights how changes in excitation--inhibition balance and metaplasticity can govern functional recovery in deprived neural systems.

\section{Introduction}

Amblyopia is among the most common causes of monocular visual impairment in children and arises from abnormal visual experience during development \cite{Birch2013,HolmesLevi2018}. Monocular deprivation (MD), strabismus, and anisometropia disrupt balanced binocular drive and can produce persistent reductions in cortical responsiveness to the affected eye \cite{Birch2013,Rittenhouse1999,FrenkelBear2004}. Although treatments such as patching or pharmacological penalization can improve vision during childhood, treatment efficacy generally declines with age as visual cortical plasticity becomes more constrained beyond the critical period \cite{HolmesLevi2018,Hensch2005}.

Experimental studies nevertheless demonstrate that substantial recovery can be induced beyond the critical period under appropriate conditions \cite{Fong2016,Fong2021,LeetBearGaier2022}. In particular, transient retinal inactivation with tetrodotoxin (TTX) can produce rapid and durable recovery of amblyopic-eye function following long-term MD \cite{Fong2016,Fong2021}. Direct comparison with reverse occlusion (RO) shows that fellow-eye inactivation produces stronger and more persistent recovery at ages when RO is ineffective or transient \cite{Fong2021}. Recent work further identifies retinal-inactivation-induced burst firing in the dorsal lateral geniculate nucleus (dLGN) as a critical trigger for recovery \cite{EchavarriLeet2025}. These findings sharpen the mechanistic question addressed here: how are changes initiated by fellow-eye inactivation translated into a cortical dynamical state that is more permissive for recovery than that produced by RO?

Several theoretical frameworks provide complementary accounts of experience-dependent cortical plasticity. Hebbian and bidirectional synaptic-plasticity models emphasize activity-dependent strengthening and weakening of synapses \cite{KirkwoodBear1995,Rittenhouse1999,Bear2003}. Homeostatic plasticity provides compensatory mechanisms that stabilize neuronal and circuit activity in the face of persistent changes in input \cite{Turrigiano1998,TurrigianoNelson2004}. BCM-type theories combine activity-dependent synaptic modification with a sliding modification threshold that regulates the balance between potentiation and depression according to recent postsynaptic activity \cite{BienenstockCooperMunro1982,Bear2003,UdeigweMunroErmentrout2017,Udeigwe2024}. Experimental studies in visual cortex further support experience-dependent metaplasticity and demonstrate that the conditions for potentiation and depression depend on prior cortical activity \cite{PhilpotEspinosaBear2003,LeetBearGaier2022}. However, how these plasticity mechanisms interact with cortical excitation--inhibition balance to distinguish recovery under TTX from RO remains incompletely understood.

A major challenge is that recovery depends on interactions across multiple dynamical processes. Sensory deprivation alters excitatory and inhibitory activity, while intracortical inhibition constrains the expression of synaptic plasticity and activity history modifies the threshold for subsequent potentiation or depression \cite{KirkwoodBear1995,Hensch2005,PhilpotEspinosaBear2003,LeetBearGaier2022}. Consequently, interventions that appear similar at the level of interocular competition may place the cortical circuit into distinct dynamical states.

Here we propose a minimal excitatory--inhibitory circuit model that explains differential recovery from amblyopia under TTX and RO through the interaction of excitation--inhibition balance, homeostatic regulation, and BCM-like metaplasticity. The model combines fast excitatory--inhibitory population dynamics with eye-specific synaptic plasticity and adaptive modification thresholds. Using phase-plane analysis, operating-regime characterization, functional recovery measurements, and parameter robustness analysis, we show that TTX and RO drive the cortex into distinct dynamical regimes despite both reducing competition from the fellow eye.

Our central result is that fellow-eye inactivation creates a larger plasticity-permissive operating regime characterized by reduced inhibition and a larger positive separation between excitatory activity and the BCM modification threshold. This state promotes stronger recovery of amblyopic-eye function and greater correction of ocular dominance than RO. More broadly, the model suggests that recovery from deprivation may be governed not simply by sensory input itself, but by the dynamical state of the cortical circuit through which that input is processed.

\section{Methods}

\subsection{Model overview}
\begin{figure}[h!]
\centering
\includegraphics[width=0.8\textwidth]{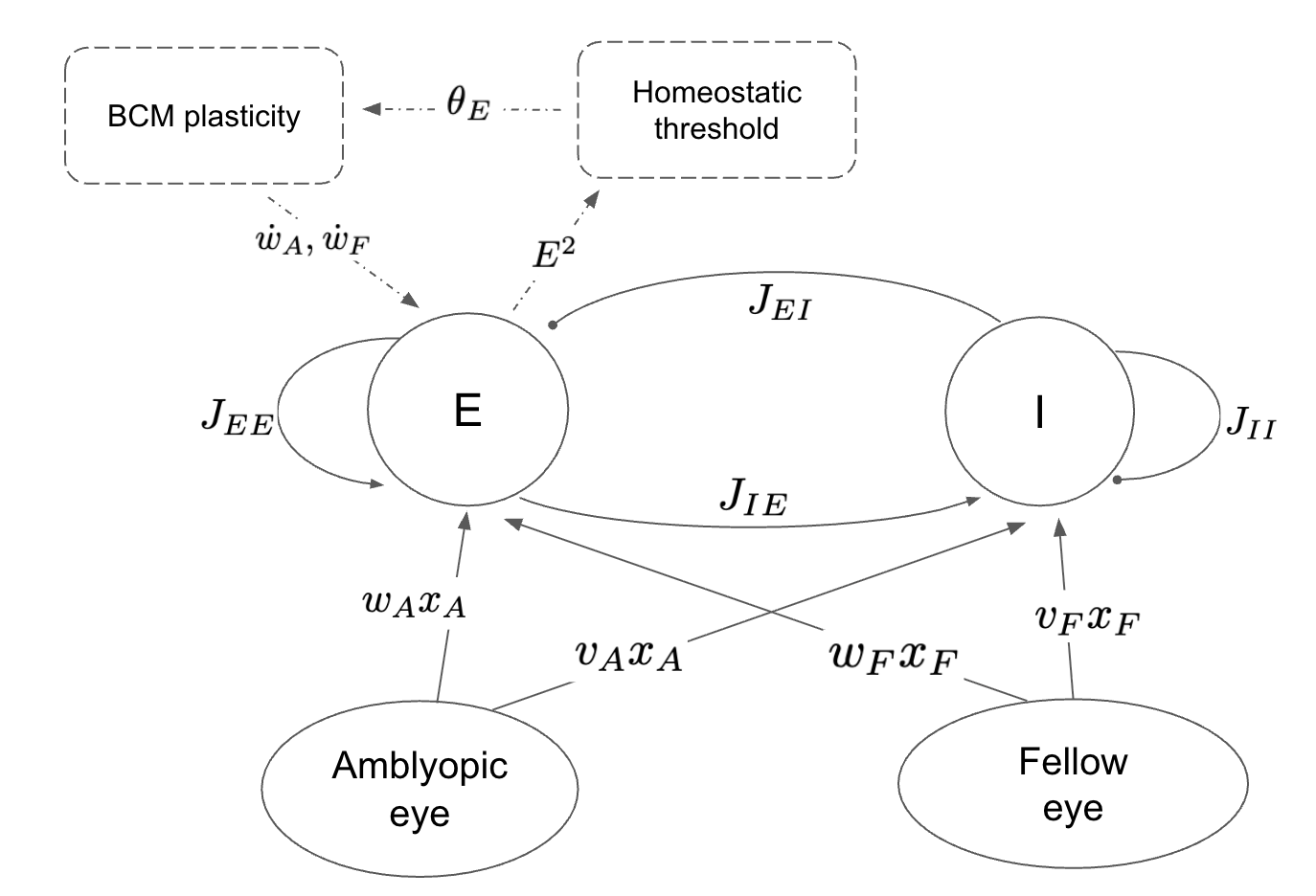}
\caption{ 
Minimal excitatory/inhibitory circuit schematic. Amblyopic and fellow-eye inputs provide bottom-up drive to excitatory and inhibitory populations. Eye-specific excitatory synapses undergo BCM-type plasticity, while homeostatic adaptation regulates the sliding plasticity threshold through recent excitatory activity.
}
\label{fig:minimal_ei_circuit},
\end{figure}

We consider a minimal cortical microcircuit consisting of one excitatory population and one inhibitory population receiving eye-specific inputs from the amblyopic and fellow eyes  \cite{WilsonCowan1972}. The excitatory population represents aggregate cortical activity associated with binocular visual processing, while the inhibitory population provides recurrent suppression and gain control. As shown in Figure \ref{fig:minimal_ei_circuit}, the model combines fast excitatory–inhibitory population dynamics, homeostatic adaptation of excitatory gain, eye-specific feedforward inputs, and BCM-like synaptic plasticity.

\subsubsection*{Fast excitatory--inhibitory dynamics}
The fast subsystem evolves according to:

\begin{align}
\tau_E \frac{dE}{dt} &= -E + \phi_E\left(g_E(w_A x_A + w_F x_F + J_{EE}E - J_{EI}I) + b_E\right),
\label{eq:ei_system_e}
\\
\tau_I \frac{dI}{dt} &= -I + \phi_I\left(v_A x_A + v_F x_F + J_{IE}E - J_{II}I + b_I\right).
\label{eq:ei_system_i}
\end{align}

Here, $E$ and $I$ denote excitatory and inhibitory population activity, respectively. The constants $\tau_E$ and $\tau_I$ are the characteristic time constants governing the rates at which the excitatory and inhibitory populations relax toward their instantaneous firing-rate responses. The quantities $x_A$ and $x_F$ represent amblyopic-eye and fellow-eye inputs, while $w_A$ and $w_F$ are plastic feedforward synaptic weights onto the excitatory population. Recurrent coupling strengths are denoted by $J_{ab}$, where $b$ identifies the presynaptic (sending) population and $a$ the postsynaptic (receiving) population. The parameter $g_E$ is a slowly varying excitatory gain term, while $v_A$ and $v_F$ determine eye-specific feedforward recruitment of the inhibitory population. Finally, $\phi_E$ and $\phi_I$ denote nonlinear activation functions, implemented here as rectified linear (ReLU) or softplus nonlinearities.

\subsubsection*{Eye-Specific Feedforward Input}

The variables $x_A(t)$ and $x_F(t)$ represent effective visual drive from
the amblyopic and fellow eyes, respectively. Their magnitudes are specified
by the experimental condition: comparable input from both eyes during
baseline, reduced amblyopic-eye input during MD, residual fellow-eye input
during RO, and complete elimination of fellow-eye input during TTX.
Condition-specific values are listed in the Appendix. A small stochastic
component is added to the residual fellow-eye input only in the RO
operating-regime analysis, as described in Section~\ref{subsection: operating-regime}.

\subsubsection*{Homeostatic Excitatory Gain Regulation}
To maintain stable firing rates over long timescales, the model includes a slow homeostatic regulation of excitatory gain  \cite{Turrigiano1998,TurrigianoNelson2004,Desai2002,MrsicFlogel2007,Keck2013}. To capture this, the excitatory gain variable $g_E$ evolves according to
\begin{equation}
\tau_g
\frac{dg_E}{dt}
=
g_E(E^\ast-E),
\end{equation}

where $E^\ast$ denotes a target excitatory activity level.
When excitatory activity fell below the target,
$g_E$ increases, amplifying subsequent sensory inputs.
When activity exceeded the target,
$g_E$ decreases.
This multiplicative homeostatic mechanism provides a slow
activity-normalization process that interacts with BCM
plasticity during deprivation and recovery.

\subsubsection*{BCM-like plasticity}
Excitatory feedforward synapses evolve according to BCM-like plasticity \cite{BienenstockCooperMunro1982, BearCooperEbner1987, Blais2008,CooperBear2012,UdeigweMunroErmentrout2017,Udeigwe2024}:
\begin{align}
\frac{dw_A}{dt} &= \eta_A x_A E(E-\theta_E) - \lambda_w w_A,
\label{eq:bcm_a}
\\
\frac{dw_F}{dt} &= \eta_F x_F E(E-\theta_E) - \lambda_w w_F.
\label{eq:bcm_f}
\end{align}
Here, $\theta_E$ denotes a sliding plasticity threshold and $\eta_A,\eta_F$ are learning rates and and $\lambda_w$ is a passive synaptic decay rate. The threshold evolves according to:
\begin{equation}
\tau_\theta \frac{d\theta_E}{dt} = E^2 - \theta_E.
\label{eq:bcm_threshold}
\end{equation}
This formulation allows the modification threshold to adapt to long-term activity levels.

\subsection{Experimental protocol}

Each simulation consists of four sequential phases:

\begin{itemize}
    \item \textbf{Baseline binocular vision:} Both eyes receive comparable visual input ($x_A \approx x_F$), establishing the homeostatic binocular operating state.

    \item \textbf{Monocular deprivation (MD):} Input from the amblyopic eye is substantially reduced relative to the fellow eye ($x_A \ll x_F$), producing deprivation-induced cortical plasticity.

    \item \textbf{Recovery intervention:} Recovery proceeds using one of two protocols \cite{Fong2016,Fong2021}:
    \begin{itemize}
        \item \textbf{TTX:} The fellow eye is completely silenced ($x_F \approx 0$).
        \item \textbf{RO:} The fellow eye retains weak residual activity ($0 < x_F < x_A$).
    \end{itemize}

    \item \textbf{Post-recovery binocular vision:} Normal binocular input is restored to both eyes ($x_A \approx x_F$) to evaluate the persistence of functional recovery.
\end{itemize}

\begin{figure}[h!]
\centering
\includegraphics[width=\textwidth]{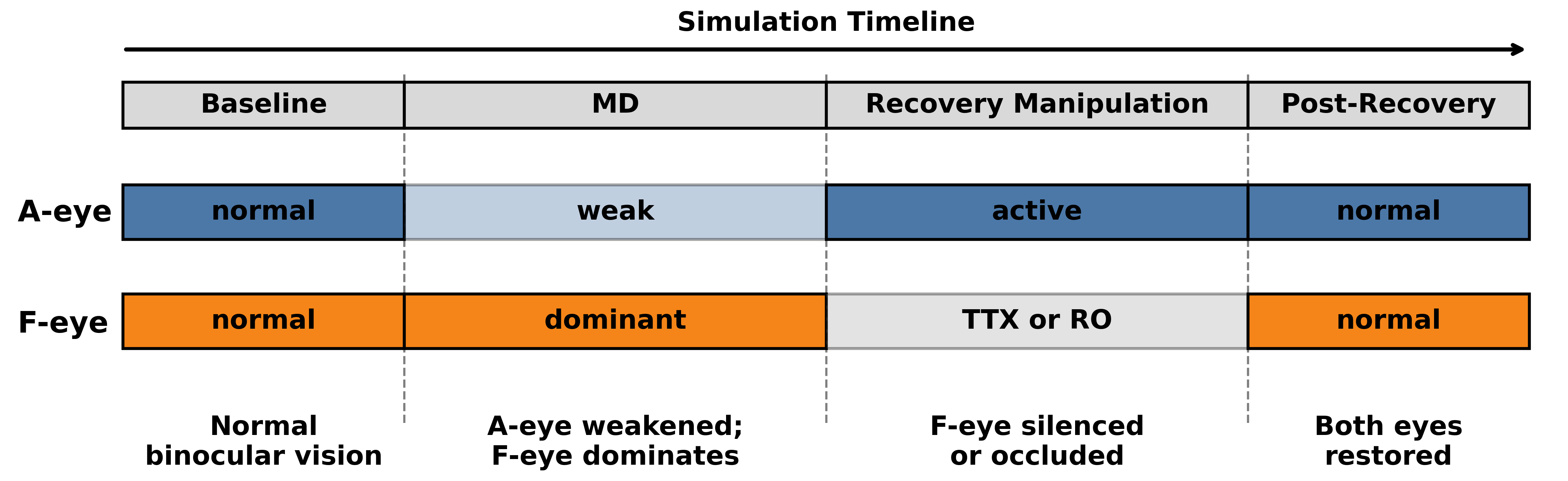}
\caption{
Simulation timeline showing baseline, monocular deprivation, TTX or reverse occlusion intervention, and post-recovery phases.
}
\label{fig:timeline}
\end{figure}

This protocol is summarized in Figure \ref{fig:timeline}. Because TTX and RO differ only in the level of residual fellow-eye activity during the recovery phase, this protocol provides a direct quantitative comparison of the two interventions under identical deprivation histories.

\subsection{Ocular dominance index}

To quantify eye dominance, we compute the ocular dominance index (ODI) \cite{Cang2005},

\begin{equation}
\mathrm{ODI}
=
\frac{R_F-R_A}{R_F+R_A},
\end{equation}

where $R_A$ and $R_F$ denote the steady-state excitatory responses elicited by isolated stimulation of the amblyopic and fellow eyes, respectively. Positive ODI values indicate fellow-eye dominance, whereas values approaching zero reflect balanced binocular responses.

\section{Results}

\subsection{Reduced Phase-Plane and BCM Analysis}

To gain mechanistic insight into how fellow-eye silencing alters cortical plasticity, we analyzed the fast excitatory--inhibitory (E/I) subsystem (Equations~\ref{eq:ei_system_e}--\ref{eq:ei_system_i}). Synaptic weights and metaplastic variables were held fixed, allowing the neuronal dynamics to rapidly converge to condition-dependent steady states. Complete parameter values for this section and subsequent sections are listed in the Appendix.

\begin{figure}[h!]
\centering
\includegraphics[width=\textwidth]{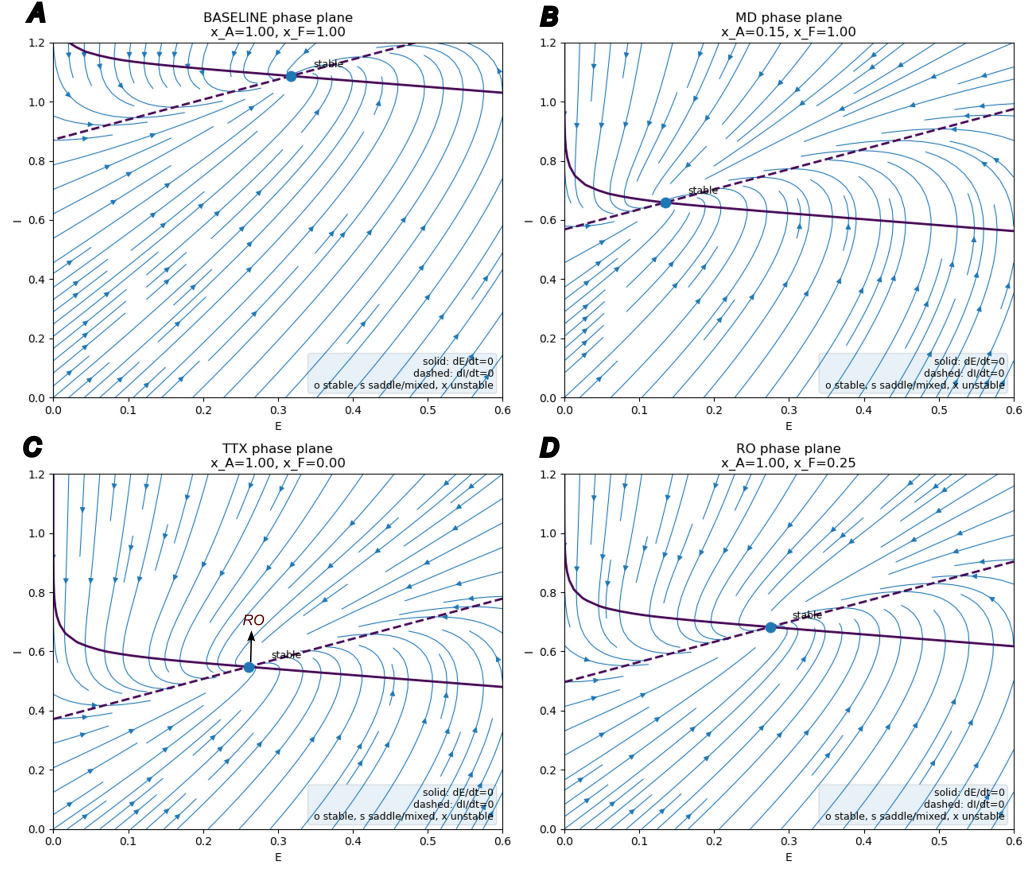}
\caption{ 
Reduced phase-plane analysis of the fast E/I subsystem under four visual conditions. 
(A) Baseline binocular vision produced a stable moderate-activity fixed point. 
(B) Monocular deprivation (MD) shifted the operating point toward reduced excitatory and inhibitory activity due to suppression of amblyopic-eye drive. 
(C) Fellow-eye tetrodotoxin silencing (TTX) reduced inhibitory activity while preserving moderate excitatory activity, producing a disinhibited recovery-permissive regime. 
(D) Reverse occlusion (RO) also restored excitatory activity but maintained stronger inhibition than TTX because residual fellow-eye activity remained present. 
Solid curves denote the $dE/dt=0$ nullcline and dashed curves denote the $dI/dt=0$ nullcline. Filled circles indicate stable fixed points.
}
\label{fig:phase_plane}
\end{figure}

For each visual condition (baseline, MD, TTX, and RO), the reduced system was numerically equilibrated to obtain the fixed points shown in Figure~\ref{fig:phase_plane}.

The corresponding equilibria were
\begin{equation}
\begin{aligned}
\text{Baseline:}\quad &(E^\ast,I^\ast)=(0.3172,\,1.0867),\\
\text{MD:}\quad &(E^\ast,I^\ast)=(0.1345,\,0.6591),\\
\text{TTX:}\quad &(E^\ast,I^\ast)=(0.2605,\,0.5482),\\
\text{RO:}\quad &(E^\ast,I^\ast)=(0.2746,\,0.6828).
\end{aligned}
\end{equation}

The phase-plane geometry reveals three key features. First, MD shifts the operating point toward substantially lower excitatory activity, reflecting reduced amblyopic-eye drive. Second, TTX produces the lowest inhibitory activity while maintaining moderate excitation, placing the network in a disinhibited operating regime. Third, RO restores excitation to a similar level but preserves substantially stronger inhibition because residual fellow-eye activity remains present.

To relate these operating points to synaptic plasticity, we define an amblyopic-eye plasticity score based on the classical BCM learning rule (Equation~\ref{eq:bcm_a}),
\begin{equation}
S_A
=
\eta_Ax_AE^\ast(E^\ast-\theta_E)
-
\lambda_ww_A,
\end{equation}
where $E^\ast$ is the excitatory fixed point. Since synaptic weights remain fixed in the reduced analysis, $S_A$ serves only as a diagnostic measure of the local propensity for amblyopic-eye potentiation.

The BCM modification threshold is approximated by
\begin{equation}
\theta_E=\alpha(E^\ast)^2,
\end{equation}
with $\alpha=2$, compensating for the fact that all reduced equilibria satisfy $E^\ast<1$, for which the unscaled quadratic threshold would underestimate the modification threshold.

Setting $\eta_A=1$, $\lambda_w=0.01$, and $w_A=0.5$ yields
\begin{equation}
S_A
=
x_AE^\ast(E^\ast-\theta_E)-0.005,
\end{equation}
whose sign distinguishes potentiation ($S_A>0$) from net weakening ($S_A<0$).

For the TTX protocol,
\begin{equation}
\begin{aligned}
\text{Baseline:}\quad &S_A=0.0313,\\
\text{MD:}\quad &S_A=-0.0035,\\
\text{TTX:}\quad &S_A=0.0284,\\
\text{Post:}\quad &S_A=0.0316,
\end{aligned}
\end{equation}

whereas for the RO protocol,
\begin{equation}
\begin{aligned}
\text{Baseline:}\quad &S_A=0.0313,\\
\text{MD:}\quad &S_A=-0.0035,\\
\text{RO:}\quad &S_A=0.0302,\\
\text{Post:}\quad &S_A=0.0315.
\end{aligned}
\end{equation}

These results show that MD shifts the amblyopic pathway into a net weakening regime, whereas both TTX and RO restore a positive plasticity regime. Although the two interventions produce similar BCM plasticity scores, the phase-plane analysis demonstrates that they occupy distinct E/I operating regimes, with TTX maintaining substantially lower inhibition than RO. Thus, the plasticity score identifies whether the network is permissive for potentiation, while the phase-plane analysis reveals the inhibitory context in which that plasticity occurs.

This reduced analysis serves as a mechanistic approximation of the full dynamical model. In the full simulations, synaptic weights, inhibitory recruitment, and the BCM threshold evolve jointly over time. Nevertheless, the reduced analysis explains how fellow-eye inactivation shifts the amblyopic pathway from net weakening to a plasticity-permissive regime through coupled excitatory--inhibitory rebalancing and metaplasticity.

\subsection{Distinct operating regimes emerge under baseline, monocular deprivation, reverse occlusion, and fellow-eye inactivation}
\label{subsection: operating-regime}
\begin{figure}[h!]
\centering
\includegraphics[width=\textwidth]{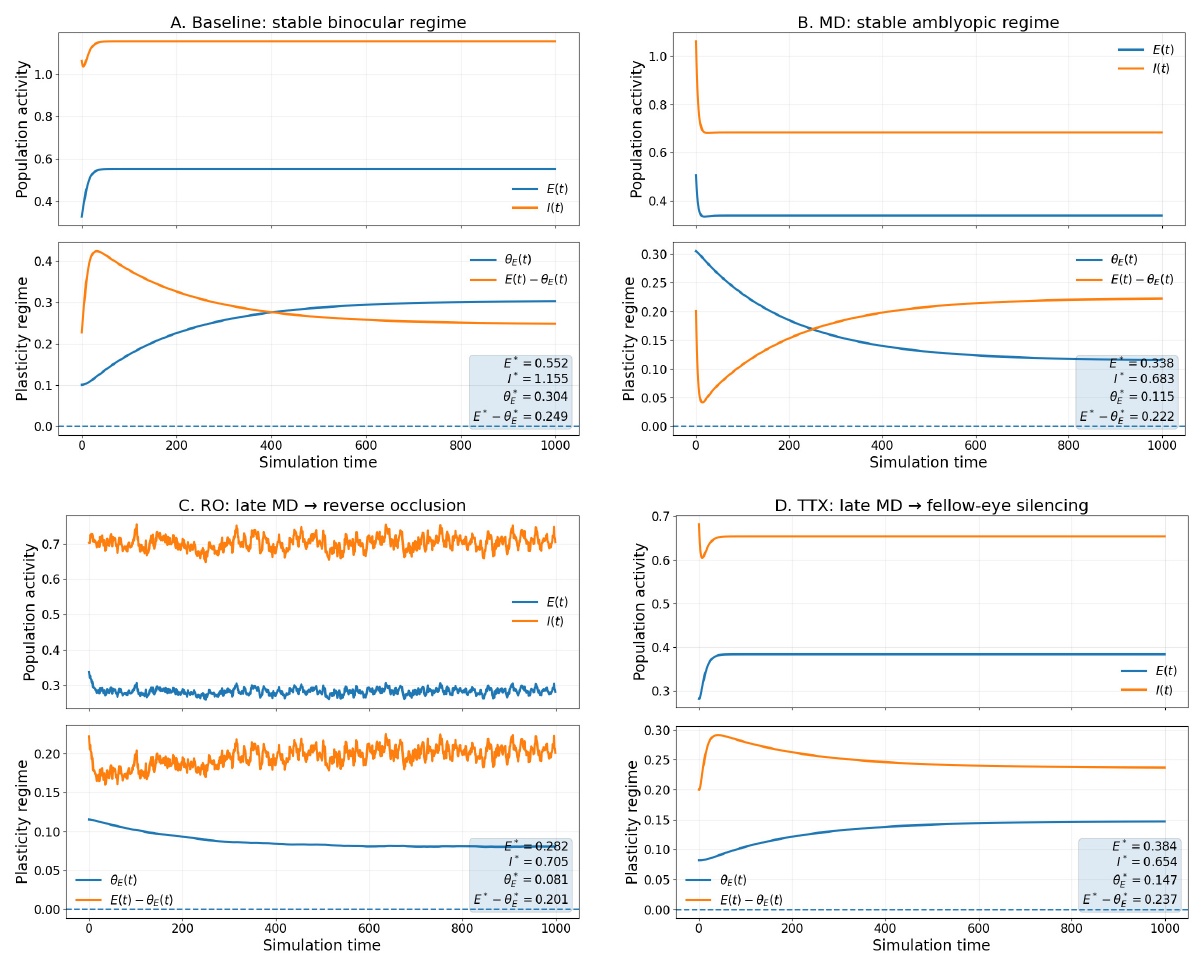}
\caption{ 
Operating regimes of the excitatory--inhibitory circuit under baseline binocular vision, monocular deprivation (MD), reverse occlusion (RO), and fellow-eye tetrodotoxin (TTX) inactivation. For each condition, the upper panel shows excitatory ($E$) and inhibitory ($I$) activity, while the lower panel shows the BCM modification threshold ($\theta_E$) and the plasticity-driving quantity $E-\theta_E$. Baseline establishes a balanced binocular regime, MD shifts the network to a lower-activity amblyopic regime, RO partially restores cortical activity while preserving inhibition, and TTX produces the lowest inhibition and the largest plasticity-permissive window among the recovery conditions.
}
\label{fig:operating_regimes}
\end{figure}

To determine how different visual manipulations alter cortical plasticity, we examined the operating regimes of the E/I system under baseline binocular vision, monocular deprivation (MD), reverse occlusion (RO), and fellow-eye tetrodotoxin (TTX) inactivation (Figure~\ref{fig:operating_regimes}). For each condition, we monitored excitatory activity ($E$), inhibitory activity ($I$), the BCM modification threshold ($\theta_E$), and the plasticity-driving quantity $E-\theta_E$. Synaptic weights were held fixed to isolate the effects of visual input on the fast E/I dynamics and the associated BCM threshold. Plastic changes in $w_A$ and $w_F$ are examined in the subsequent recovery simulations.

\subsubsection*{Baseline binocular vision}
Under normal binocular stimulation ($x_A=x_F=1$), the network converged to a stable operating regime with moderate excitatory activity ($E^\ast=0.5523$), stronger inhibitory activity ($I^\ast=1.1550$), and a BCM threshold of $\theta_E^\ast=0.3036$ (Figure~\ref{fig:operating_regimes}A). Excitatory activity remained above threshold ($E^\ast-\theta_E^\ast=0.2487$), establishing the balanced reference state against which all subsequent manipulations were compared.

\subsubsection*{Monocular deprivation}
Reducing amblyopic-eye input ($x_A\approx0.15,\;x_F=1$) shifted the circuit to a lower-activity regime (Figure~\ref{fig:operating_regimes}B). Both excitation and inhibition decreased ($E^\ast=0.3378$, $I^\ast=0.6833$), and the BCM threshold fell to $\theta_E^\ast=0.1154$. Although $E^\ast-\theta_E^\ast$ remained positive ($0.2224$), the weak amblyopic input limited the expression of plasticity through the deprived pathway.

\subsubsection*{Reverse occlusion}
During RO, the amblyopic-eye input was fixed at $x_A=1$, while the residual fellow-eye input included a small Gaussian noise component to represent ongoing activity despite occlusion,
\[
x_F(t)=\max\!\left[0,\,G_F+\sigma_F\xi_F(t)\right],
\]
with $G_F=0.30$ and $\sigma_F=0.10$. Here, $\xi_F(t)$ denotes a zero-mean, unit-variance Gaussian noise process, and rectification prevents negative visual drive.

This condition ($x_F\approx0.29$) reduced interocular competition while preserving substantial inhibitory activity (Figure~\ref{fig:operating_regimes}C). The network settled at $E^\ast=0.2852$ and $I^\ast=0.7167$, with $\theta_E^\ast=0.0820$ and $E^\ast-\theta_E^\ast=0.2032$. Thus, RO restored a plasticity-permissive regime while maintaining stronger inhibition than TTX.

\subsubsection*{Fellow-eye inactivation}
Complete fellow-eye silencing ($x_A=1,\;x_F=0$) produced the most permissive recovery regime (Figure~\ref{fig:operating_regimes}D). Inhibitory activity decreased to $I^\ast=0.6540$, the lowest among the recovery conditions, while excitatory activity remained relatively elevated ($E^\ast=0.3840$). Consequently, the BCM threshold stabilized at $\theta_E^\ast=0.1470$, yielding a larger plasticity signal ($E^\ast-\theta_E^\ast=0.2370$) than observed under RO.

\subsubsection*{Remark: TTX creates a distinct plasticity regime}
Although both RO and TTX reduce competition from the fellow eye, they drive the cortex into distinct operating regimes. Relative to RO, TTX produces lower inhibitory activity while maintaining a larger separation between excitatory activity and the BCM threshold. This enlarged plasticity-permissive window provides a mechanistic explanation for the stronger recovery observed following fellow-eye inactivation.

\subsection{Fellow-eye inactivation produces stronger functional recovery than reverse occlusion}
\begin{figure}[h!]
\centering
\includegraphics[width=\textwidth]{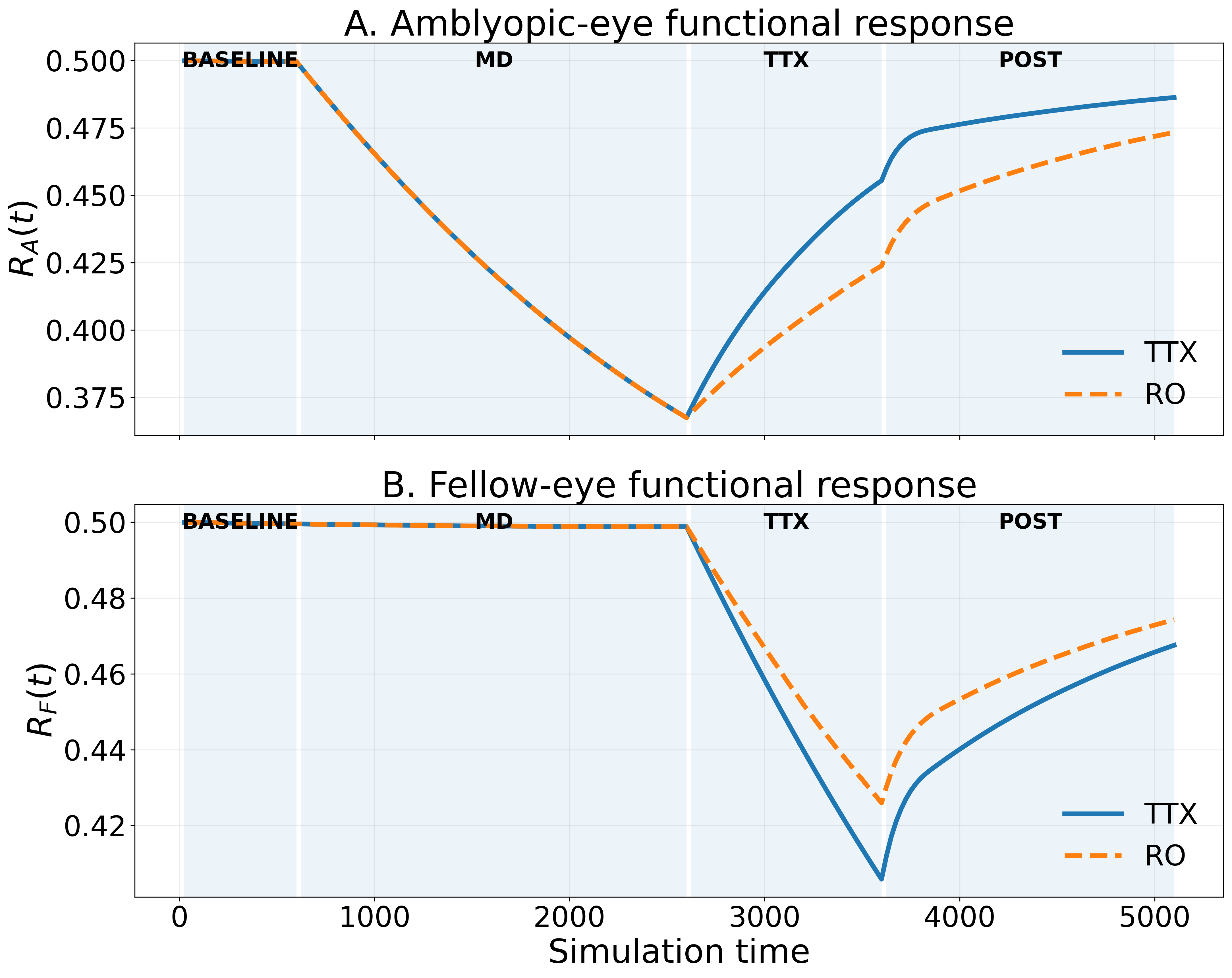}
\caption{ \textbf{Fellow-eye inactivation produces stronger functional recovery than reverse occlusion.}
Monocular functional responses were measured by probing the evolving cortical circuit separately through the amblyopic and fellow eyes, with synaptic weights and circuit parameters held fixed during each probe. \textbf{(A)} Under baseline binocular vision, amblyopic-eye responsiveness ($R_A$) is stable. Monocular deprivation (MD) progressively reduces $R_A$, producing a pronounced functional deficit in the deprived pathway. During recovery, fellow-eye tetrodotoxin (TTX) inactivation produces a larger increase in $R_A$ than reverse occlusion (RO), and this advantage persists following restoration of binocular vision. \textbf{(B)} Fellow-eye responsiveness ($R_F$) remains largely preserved during MD but decreases during both recovery interventions. The reduction is greater under TTX than RO, consistent with stronger suppression of the previously dominant fellow-eye pathway. Together, these results show that TTX produces a larger redistribution of functional responsiveness toward the amblyopic eye than RO, characterized by stronger recovery of deprived-eye function and greater suppression of fellow-eye dominance.}
\label{fig:functional_recovery}
\end{figure}

\begin{figure}[h!]
\centering
\includegraphics[width=\textwidth]{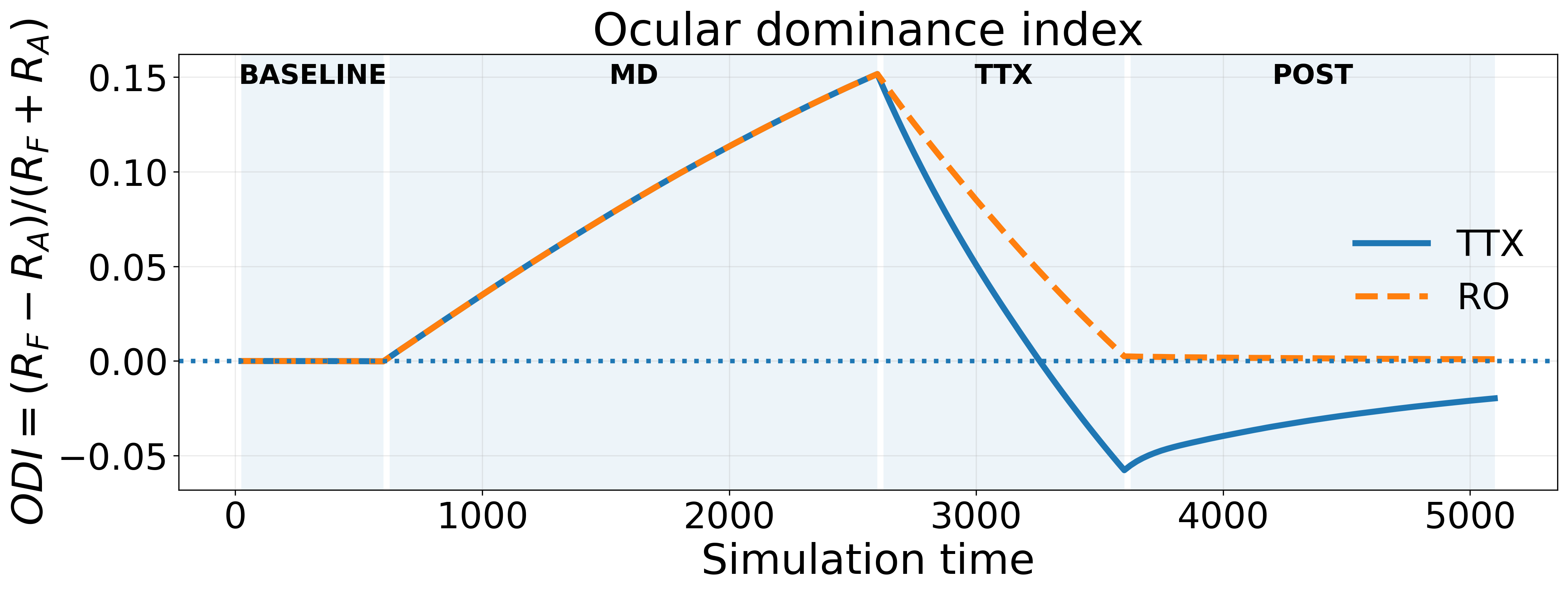}
\caption{ \textbf{Fellow-eye inactivation produces a larger correction of ocular dominance than reverse occlusion.}
MD shifts the ocular dominance index (ODI) from balanced binocular responses toward fellow-eye dominance. Both recovery interventions reduce this imbalance, but TTX produces a larger shift than RO, transiently reversing ocular dominance toward the amblyopic eye. Following restoration of binocular vision, ODI gradually returns toward balance, with the TTX-induced shift remaining evident during the post-treatment period.
}
\label{fig:odi}
\end{figure}

To determine whether the distinct excitatory--inhibitory operating regimes identified above translate into differential recovery of visual function, we next enabled eye-specific synaptic plasticity and quantified the resulting monocular functional responses and ocular dominance dynamics (Figures~\ref{fig:functional_recovery} and \ref{fig:odi}). In contrast to the operating-regime analysis, where synaptic weights were held fixed, these simulations allowed the amblyopic-eye and fellow-eye synaptic weights ($w_A$ and $w_F$) to evolve according to a modified BCM learning rule.

\subsubsection*{A baseline-referenced BCM rule}
In the mature visual cortex, normal binocular vision represents a stable homeostatic operating state \cite{Desai2002,MrsicFlogel2007,Keck2013,TurrigianoNelson2004}. Accordingly, we modified the classical BCM learning rule so that synaptic plasticity is driven by deviations from this calibrated baseline rather than by absolute levels of cortical activity. This formulation preserves a stable binocular reference state while allowing visual perturbations to induce activity-dependent plasticity.

The classical BCM-inspired weight update (Equations~\ref{eq:bcm_a}--\ref{eq:bcm_f})
produced progressive decay of both eye-specific synapses even under normal binocular stimulation, preventing the model from maintaining a stable adult binocular state. To eliminate this behavior, we introduced a baseline-referenced BCM signal.

We first define the instantaneous modification signal,
\begin{equation}
B(E,\theta_E) = E(E-\theta_E),
\end{equation}
and its value at the calibrated binocular operating point,
\begin{equation}
B_0= E_{\mathrm{base}}\left(E_{\mathrm{base}}-\theta_{E,\mathrm{base}}\right),
\end{equation}
where $E_{\mathrm{base}}$ and $\theta_{E,\mathrm{base}}$ denote the steady-state excitatory activity and BCM threshold under normal binocular vision. Plasticity is then driven by the deviation
\begin{equation}
\Delta B = B(E,\theta_E)-B_0.
\end{equation}

The eye-specific feedforward weights evolve according to
\begin{align}
\frac{dw_A}{dt}&=\eta_A x_A \Delta B-\lambda_w \left(w_A-w_{\mathrm{home}}\right),
\label{eq:wa_rule}
\\
\frac{dw_F}{dt}&=\eta_F x_F \Delta B-\lambda_w\left(w_F-w_{\mathrm{home}}\right),
\label{eq:wf_rule}
\end{align}
where $w_{\mathrm{home}}$ denotes the homeostatic binocular synaptic strength. 

Under normal binocular stimulation,
\begin{equation}
E \approx E_{\mathrm{base}},
\qquad
\theta_E \approx \theta_{E,\mathrm{base}},
\qquad
w_i \approx w_{\mathrm{home}},
\end{equation}
so that $\Delta B \approx 0$ and the synaptic weights remain approximately constant. Consequently, the baseline-referenced BCM rule preserves a stable binocular operating point while allowing eye-specific visual perturbations to drive adaptive synaptic plasticity.

\subsubsection*{Monocular deprivation selectively impairs amblyopic-eye responsiveness}
To quantify recovery at the functional level, we measured monocular probe responses rather than directly examining synaptic weights. At each time point, the learned circuit was probed separately through the amblyopic eye and the fellow eye while all synaptic and circuit parameters were held fixed. The resulting steady-state excitatory responses were denoted $R_A$ and $R_F$, respectively.

Under baseline binocular vision (Figure~\ref{fig:functional_recovery}), the two eyes generated nearly identical responses,
\begin{equation}
R_A \approx R_F \approx 0.50,
\end{equation}
yielding an ocular dominance index near zero,
\begin{equation}
ODI
=
\frac{R_F-R_A}{R_F+R_A}
\approx 0.
\label{eq:odi}
\end{equation}

Following monocular deprivation, the amblyopic-eye response declined substantially, decreasing from approximately $0.50$ to $0.37$, whereas the fellow-eye response remained largely unchanged. Consequently, the ocular dominance index increased from approximately $0$ to $0.15$, indicating the emergence of pronounced fellow-eye dominance.

These results reproduce the expected functional signature of amblyopia and demonstrate that deprivation selectively impairs responsiveness through the deprived pathway while preserving responsiveness through the non-deprived pathway \cite{WieselHubel1963,Rittenhouse1999,FrenkelBear2004}.

\subsubsection*{Fellow-eye inactivation produces stronger recovery than reverse occlusion}

Fellow-eye tetrodotoxin inactivation induced substantial recovery of the amblyopic-eye pathway. During the recovery phase, the amblyopic-eye response increased from approximately $R_A = 0.37$ to $R_A = 0.46$, while the fellow-eye response decreased from approximately $R_F = 0.50$ to $R_F = 0.41$. As a result, ocular dominance shifted from strong fellow-eye dominance ($ODI \approx 0.15$) to a transient amblyopic-eye bias ($ODI \approx -0.06$).

Reverse occlusion produced qualitatively similar but weaker effects. During RO, the amblyopic-eye response increased only to approximately $R_A = 0.42$, while the fellow-eye response decreased to approximately $R_F = 0.43$. Consequently, the ocular dominance index approached zero but did not exhibit the pronounced amblyopic-eye shift observed following TTX.

This model result is consistent with experimental observations that fellow-eye
inactivation produces stronger and more persistent recovery than reverse occlusion \cite{Fong2021}.

\subsubsection*{Fellow-eye inactivation and reverse occlusion occupy distinct regions of the functional recovery landscape}
Direct comparison of the recovery conditions demonstrates that TTX and RO produce fundamentally different patterns of cortical recovery. Although both manipulations reduced fellow-eye dominance and improved amblyopic-eye responsiveness, TTX consistently generated larger increases in $R_A$, stronger reductions in $R_F$, and a correspondingly larger shift in ocular dominance.

Importantly, these differences emerged despite both interventions acting on the fellow-eye pathway. The results therefore support the hypothesis that the superior efficacy of fellow-eye inactivation arises not simply from removal of visual competition, but from the distinct cortical operating regime induced by TTX. Consistent with the operating-regime analysis presented above, TTX generates a more plasticity-permissive state characterized by reduced inhibition, elevated modification potential, and enhanced recovery of deprived-eye function.

Following restoration of binocular vision, both recovery conditions converged toward balanced binocular responses. However, the larger amblyopic-eye response achieved during TTX recovery persisted throughout the post-treatment period, indicating a more substantial restoration of deprived-eye function than that achieved by reverse occlusion.

\subsection{TTX-mediated recovery is robust to local parameter perturbations}
\begin{figure}[h!]
\centering
\includegraphics[width=\textwidth]{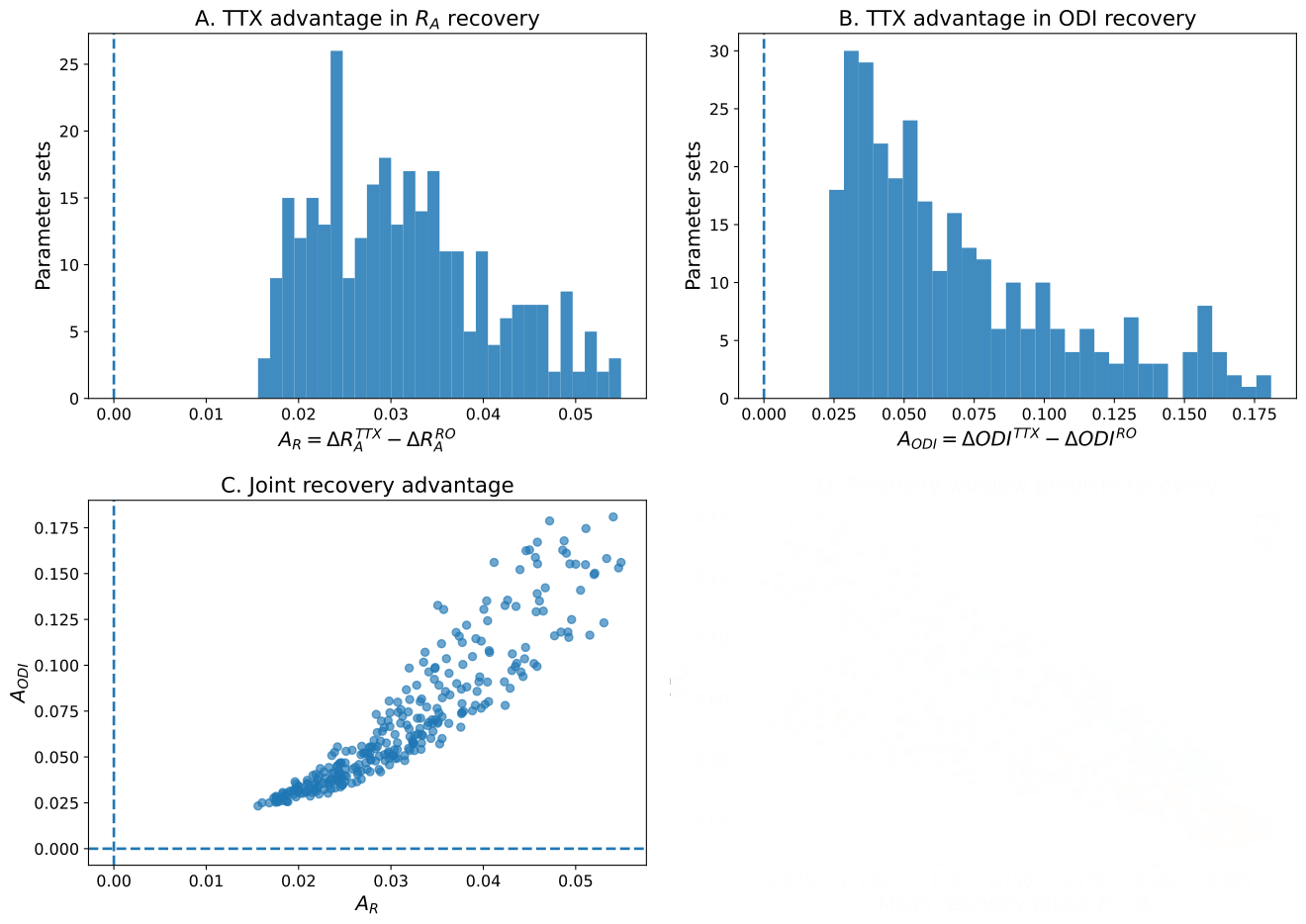}
\caption{
\textbf{TTX-mediated recovery is robust to local parameter perturbations.}
Robustness analysis over $300$ randomly perturbed parameter sets. Parameters controlling E/I coupling, BCM threshold adaptation, synaptic learning, and synaptic homeostasis were independently varied by $ \pm 25 \%$ around the representative parameter set. For each parameter set, paired TTX and RO simulations were run from the same post-MD state.
(A) Distribution of the amblyopic-eye recovery advantage
$A_R=\Delta R_A^{TTX}-\Delta R_A^{RO}.$
Positive values indicate greater amblyopic-eye recovery under TTX than RO. TTX outperformed RO in all sampled parameter sets.
(B) Distribution of the ocular-dominance recovery advantage
$A_{ODI}=\Delta ODI^{TTX}-\Delta ODI^{RO}.$
Positive values indicate stronger correction of fellow-eye dominance under TTX than RO. TTX produced greater ODI recovery in all sampled parameter sets.
(C) Joint recovery advantage. Each point corresponds to one perturbed parameter set. Points in the upper-right quadrant indicate parameter sets for which TTX outperformed RO according to both amblyopic-eye response recovery and ocular-dominance recovery.
}
\label{fig:robustness}
\end{figure}

The preceding simulations demonstrate that fellow-eye TTX produces stronger amblyopic-eye recovery than reverse occlusion for a representative parameter set. To determine whether this conclusion depended on fine tuning, we performed a local robustness analysis around the parameter regime used in the main simulations.

We randomly perturbed six parameters controlling excitatory--inhibitory coupling, metaplastic adaptation, and synaptic plasticity:
\[
J_{EI},\quad J_{IE},\quad \tau_{\theta_E},\quad \eta_A,\quad \eta_F,\quad \lambda_w .
\]
For each parameter \(p\), we generated a perturbed value
\[
p^{(k)} = p_0(1+\epsilon_k),
\]
where \(p_0\) denotes the baseline value and
\[
\epsilon_k \sim \mathcal{U}(-0.25,0.25).
\]
Thus, each sampled parameter set varied the selected parameters by up to \(\pm 25\%\) relative to the values used in the representative simulation. All other parameters were held fixed. For each sampled parameter set, we ran paired simulations of the full protocol:
\[
\text{baseline} \rightarrow \text{MD} \rightarrow \text{TTX} \rightarrow \text{post},
\]
and
\[
\text{baseline} \rightarrow \text{MD} \rightarrow \text{RO} \rightarrow \text{post}.
\]
The TTX and RO simulations were initialized from the same post-MD state, ensuring that differences between recovery conditions were attributable to the recovery manipulation rather than to differences accumulated during baseline or deprivation.

For each simulation, we quantified amblyopic-eye recovery as the change in monocular amblyopic-eye response from the end of MD to the end of the recovery phase:
\[
\Delta R_A^{TTX}
=
R_{A,\mathrm{rec}}^{TTX}
-
R_{A,\mathrm{MD}},
\]
and
\[
\Delta R_A^{RO}
=
R_{A,\mathrm{rec}}^{RO}
-
R_{A,\mathrm{MD}}.
\]
We then defined the TTX recovery advantage in amblyopic-eye responsiveness as
\[
A_R
=
\Delta R_A^{TTX}
-
\Delta R_A^{RO}.
\]
Positive values of \(A_R\) indicate parameter sets for which TTX produced greater amblyopic-eye recovery than RO.

We also quantified recovery at the level of ocular dominance. Since monocular deprivation shifts the ocular dominance index toward positive values indicating fellow-eye dominance, recovery was measured as the reduction in ODI from the end of MD to the end of the recovery phase:
\[
\Delta ODI^{TTX}
=
ODI_{\mathrm{MD}}
-
ODI_{\mathrm{rec}}^{TTX},
\]
and
\[
\Delta ODI^{RO}
=
ODI_{\mathrm{MD}}
-
ODI_{\mathrm{rec}}^{RO}.
\]
The corresponding TTX advantage in ocular-dominance recovery was defined as
\[
A_{ODI}
=
\Delta ODI^{TTX}
-
\Delta ODI^{RO}.
\]
Thus, \(A_{ODI}>0\) indicates stronger correction of fellow-eye dominance under TTX than under RO.

Across \(300\) randomly perturbed parameter sets, TTX produced greater amblyopic-eye response recovery than RO in all sampled cases (Figure~\ref{fig:robustness}A). The median recovery advantage was
\[
\mathrm{median}(A_R)=0.0303,
\]
indicating that the effect was not merely a marginal numerical difference. Similarly, TTX produced greater ocular-dominance recovery than RO in all sampled parameter sets (Figure~\ref{fig:robustness}B), with
\[
\mathrm{median}(A_{ODI})=0.0574.
\]
Thus, within the local parameter neighborhood tested here, the superiority of TTX over RO was robust to moderate perturbations in E/I coupling, BCM threshold dynamics, learning rate, and synaptic homeostasis.

Finally, we examined whether the two recovery metrics were mutually consistent across parameter sets. Plotting \(A_R\) against \(A_{ODI}\) showed that the sampled points lay in the upper-right quadrant, where both measures favor TTX (Figure~\ref{fig:robustness}C). This indicates that parameter sets producing stronger amblyopic-eye response recovery under TTX also produced stronger correction of ocular dominance. Together, these analyses demonstrate that the differential recovery observed in the representative simulations was not an isolated consequence of a single finely tuned parameter set, but instead reflected a locally robust property of the modeled cortical recovery mechanism.

\section{Discussion }
\begin{figure}[]
\centering
\includegraphics[width=0.9\textwidth]{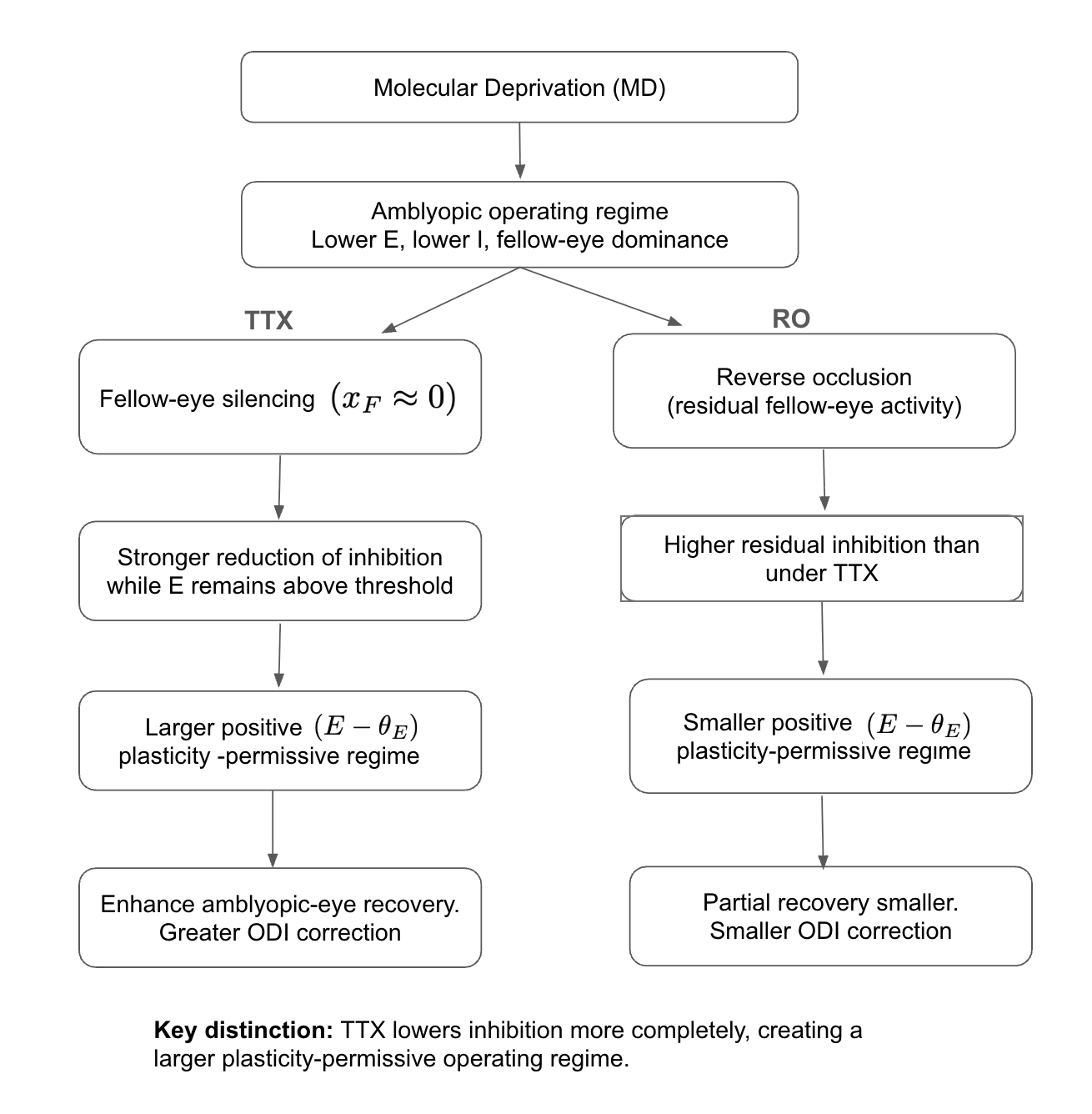}
\caption{\textbf{Proposed mechanism underlying the superior recovery produced by fellow-eye inactivation.}
Monocular deprivation (MD) shifts the circuit into an amblyopic operating regime with reduced activity and fellow-eye dominance. From this common state, TTX and reverse occlusion (RO) produce distinct recovery regimes. Fellow-eye TTX more strongly reduces inhibition while preserving sufficient excitatory activity for plasticity, generating a larger positive $E-\theta_E$ and a more plasticity-permissive state. This promotes stronger amblyopic-eye recovery and greater correction of ocular dominance. In contrast, residual fellow-eye activity under RO maintains greater inhibition, producing a smaller $E-\theta_E$, a narrower plasticity-permissive regime, and weaker recovery. These results suggest that TTX outperforms RO by shifting the circuit into a dynamical state more favorable for synaptic strengthening, rather than simply by removing interocular competition.}

\label{fig:mechanism}
\end{figure}

The analyses presented above suggest a unified mechanistic explanation for the superior recovery produced by fellow-eye tetrodotoxin (TTX) inactivation relative to reverse occlusion (RO) \cite{Fong2016,Fong2021,EchavarriLeet2025}. While both interventions reduce competition from the fellow eye, the model indicates that they alter the dynamical state of the cortical circuit in fundamentally different ways. Figure~\ref{fig:mechanism} summarizes the proposed mechanism.

Monocular deprivation (MD) first drives the network into an amblyopic operating regime characterized by reduced excitatory activity, weakened amblyopic-eye responsiveness, and a shift in ocular dominance toward the fellow eye \cite{WieselHubel1963,Rittenhouse1999,FrenkelBear2004}. Within the model, this deprived state serves as the common starting point for both recovery interventions. Thus, differences observed during recovery arise not from the deprivation phase itself, but from the distinct cortical states induced by TTX and RO.

Following fellow-eye inactivation, removal of structured fellow-eye input substantially reduces inhibitory activity while preserving sufficient excitatory drive to maintain cortical responsiveness. As shown in the operating-regime analysis (Figure~\ref{fig:operating_regimes}), this combination produces a relatively large positive separation between excitatory activity and the BCM modification threshold, $E-\theta_E$.
Because positive values of \(E-\theta_E\) favor synaptic potentiation, TTX places the network in a highly plasticity-permissive regime \cite{BienenstockCooperMunro1982,BearCooperEbner1987,CooperBear2012,UdeigweMunroErmentrout2017,Udeigwe2024}. This state supports enhanced strengthening of the amblyopic-eye pathway, leading to larger increases in amblyopic-eye responsiveness (\(R_A\)) and stronger correction of ocular dominance. This interpretation is also consistent with earlier BCM-based modeling showing that recovery from monocular deprivation depends strongly on the structure and level of activity retained in deprived and nondeprived channels \cite{Blais2008}.

Reverse occlusion produces a qualitatively different outcome. Although visual competition from the fellow eye is reduced, residual activity remains present in the previously dominant pathway. Consequently, inhibitory activity remains elevated relative to the TTX condition. The resulting increase in $E-\theta_E$
is therefore smaller, producing a narrower plasticity-permissive window. As a consequence, RO generates weaker strengthening of the amblyopic-eye pathway and smaller improvements in both functional recovery and ocular dominance.

Importantly, the robustness analysis demonstrated that the superiority of TTX over RO persists across a broad range of local parameter perturbations (Figure~\ref{fig:robustness}). Thus, the differential recovery observed in the representative simulations does not appear to depend on a narrowly tuned parameter set. Instead, it emerges as a robust consequence of the distinct operating regimes induced by the two recovery interventions.

Taken together, the results of this study support a coherent mechanistic framework for recovery following deprivation (Figure~\ref{fig:mechanism}). The phase-plane analysis demonstrated that baseline, MD, TTX, and RO occupy distinct regions of the circuit's dynamical landscape. The operating-regime analysis showed that TTX produces lower inhibitory activity and a larger positive \(E-\theta_E\) than RO. The functional recovery analysis revealed that these dynamical differences translate into greater recovery of amblyopic-eye responsiveness and stronger correction of ocular dominance. Finally, the robustness analysis established that these effects persist across a broad neighborhood of parameter values. Viewed collectively, these findings indicate that the efficacy of fellow-eye inactivation arises not merely from removal of interocular competition, but from its ability to shift the cortical circuit into a distinct plasticity-permissive operating regime characterized by reduced inhibition and enhanced capacity for synaptic strengthening.

Several limitations should be acknowledged. The model intentionally abstracts away many biological details, including cortical layers, multiple inhibitory subclasses, spatial organization, and spike-based dynamics. Consequently, the present work is not intended as a detailed biophysical model of visual cortex. Rather, its goal is to identify the minimal dynamical ingredients capable of explaining the differential recovery observed experimentally. Future studies could extend this framework to incorporate richer cortical architectures, multiple inhibitory populations, or large-scale network structure while preserving the mechanistic insights identified here.

More broadly, the results support a general view of cortical recovery in which successful interventions operate by transiently shifting neural circuits into states that are permissive for functional reorganization \cite{FagioliniHensch2000,Sale2007,MayaVetencourt2008,Harauzov2010,Kuhlman2013}. From this perspective, fellow-eye inactivation represents one example of a larger principle linking disinhibition, metaplasticity, and recovery after sensory deprivation \cite{CooperBear2012,Sale2007,Harauzov2010,Kuhlman2013,LeetBearGaier2022}. Understanding how therapeutic interventions reshape cortical operating regimes may therefore provide a useful framework for studying recovery not only in amblyopia, but across a wider range of disorders involving impaired plasticity and circuit dysfunction.

\section*{Acknowledgments}
This work is based upon work supported, in part, by the US Army Research Office and the DEVCOM US Army Research Laboratory under grant W911NF2110192.

\section*{Data Availability}
Code and simulation data will be made publicly available upon publication.

%$ \dot{w}_A , \dot{w}_F$ \\
%$ \theta_E$ \\
%$ E^2$ \\
%$J_{IE}$ \\
%$J_{EI}$ \\
%$J_{II}$ \\
%$J_{EE}$ \\
%$w_Ax_A$ \\
%$w_Fx_F$ \\
%$v_Ax_A$ \\
%$v_Fx_F$ \\

\appendix

\section{Parameters for Simulations}
\label{app:phase_plane_parameters}

\subsection{Fixed model parameters}
\begin{table}[!]
\centering
\caption{Fixed model parameters used  analyses.}
\label{tab:phase_plane_fixed_parameters}
\begin{tabular}{lll}
\hline
\textbf{Parameter} & \textbf{Value} & \textbf{Description} \\
\hline
$\tau_E$ & $10.0$ & Excitatory population time constant \\
$\tau_I$ & $5.0$ & Inhibitory population time constant \\
$b_E$ & $0.05$ & Excitatory bias \\
$b_I$ & $0.02$ & Inhibitory bias \\
$J_{EE}$ & $0.80$ & Recurrent excitation onto the excitatory population \\
$J_{EI}$ & $1.00$ & Inhibitory coupling onto the excitatory population \\
$J_{IE}$ & $0.95$ & Excitatory coupling onto the inhibitory population \\
$J_{II}$ & $0.40$ & Recurrent inhibition onto the inhibitory population \\
$w_A$ & $0.55$ & Amblyopic-eye feedforward weight onto excitation \\
$w_F$ & $0.55$ & Fellow-eye feedforward weight onto excitation \\
$v_A$ & $0.50$ & Amblyopic-eye feedforward recruitment of inhibition \\
$v_F$ & $0.70$ & Fellow-eye feedforward recruitment of inhibition \\
$g_E$ & $1.00$ & Excitatory gain \\
$\beta$ & $20.0$ & Softplus steepness parameter \\
\hline
\end{tabular}
\end{table}

\subsection{Condition-dependent effective inputs}
\begin{table}[h!]
\centering
\caption{Condition-dependent effective visual inputs used in the operating-regime analysis.}
\label{tab:phase_plane_inputs}
\begin{tabular}{lcc}
\hline
\textbf{Condition}
& \textbf{$x_A$}
& \textbf{$x_F$} \\
\hline
Baseline & $1.00$ & $1.00$ \\
MD       & $0.15$ & $1.00$ \\
TTX      & $1.00$ & $0.00$ \\
RO       & $1.00$ & $0.29$ \\
\hline
\end{tabular}
\end{table}

\subsection{Numerical settings}
\begin{table}[!]
\centering
\caption{Numerical settings for the reduced phase-plane analysis.}
\label{tab:phase_plane_numerics}
\begin{tabular}{lll}
\hline
\textbf{Setting} & \textbf{Value} & \textbf{Purpose} \\
\hline
$E$ plotting interval & $[0,\,0.6]$ & Excitatory phase-plane range \\
$I$ plotting interval & $[0,\,1.2]$ & Inhibitory phase-plane range \\
Grid resolution & $41\times41$ & Nullcline and vector-field evaluation \\
Initial guesses per axis & $8$ & Fixed-point search \\
Fixed-point tolerance & $10^{-6}$ & Deduplication of numerical roots \\
Root solver & \texttt{scipy.optimize.root} & Numerical fixed-point computation \\
\hline
\end{tabular}
\end{table}

\bibliographystyle{plain}
\bibliography{amblyopia_full.bib}

\end{document}